# Using online student focus groups in the development of new educational resources

Gian Carlo Diluvi, University of British Columbia, gian.diluvi@stat.ubc.ca

Sonja Isberg, Simon Fraser University, sonja_isberg@sfu.ca

Bruce Dunham, University of British Columbia, b.dunham@stat.ubc.ca

Nancy Heckman, University of British Columbia, nancy@stat.ubc.ca

Melissa Lee, University of British Columbia, melissa.lee@stat.ubc.ca

**Abstract**

Educational resources, such as web apps and self-directed tutorials, have become popular tools for teaching and active learning. Ideally, students—the intended users of these resources—should be involved in the resource development stage. However, in practice students often only interact with fully developed resources, when it might be too late to incorporate changes. Previous work has addressed this by involving students in the development of new resources via in-person focus groups and interviews. In these, the resource developers observe students interacting with the resource. This allows developers to incorporate their observations and students' direct feedback into further development of the resource. However, as a result of the COVID-19 pandemic, carrying out in-person focus groups became infeasible due to social distancing restrictions. Instead, online meetings and classes became ubiquitous. In this work, we describe a fully-online methodology to evaluate new resources in development. Specifically, our methodology consists of carrying out student focus groups via online video conferencing software. We assessed two educational resources for introductory statistics using our methodology and found that the online setting allowed us to obtain rich, detailed information from the students. We also found online focus groups to be more efficient: students and researchers did not need to travel and scheduling was not restricted by the availability of physical space. Our findings suggest that online focus groups are an attractive alternative to in-person focus groups for student assessment of resources in development, even now that pandemic restrictions are being eased.

**Keywords:** Online student focus groups; evaluation of educational resources; student feedback; teaching and learning statistics; teaching statistical power

## Introduction

An abundance of resources is available for teaching and learning, and new resources are being created every year. In particular, interactive resources such as web apps and self-directed learning tutorials have become increasingly popular tools. These interactive tools allow students to experiment with an activity, and to lead their own learning process. Despite the ubiquity of these resources and the amount of effort involved in their development, student feedback is rarely solicited in the development stage. At most, assessment and modifications are made on the fly, based on what the instructor might happen to observe in the classroom. In this study, we carry out an intentional process of soliciting student feedback in an online setting to aid the development process of two interactive web apps.

Interactive web apps can be a powerful tool for student learning, if set up properly. Research suggests that they are most effective if students' interaction is carefully structured (Lane & Peres, 2006), such as using activities or worksheets to help guide the interaction. As a result, each of the web apps in this study is accompanied by a structured activity sheet intended to guide the students' learning.

*Student Involvement in the Development of New Resources*

Research on approaches to evaluate new resources often breaks the evaluation process down into several steps. For example, Ooms and Garfield (2008) proposed a model that involved four components: (a) planning the evaluation process; (b) evaluation of the educational value, by observing user interaction with the resource; (c) evaluation of resource use, such as via surveys; and (d) evaluation of the educational impact, such as via student tests.

To address steps (b)-(c) of the Ooms and Garfield framework, evaluation of the resource can be conducted in several ways. Dunham et al. (2018) advise that interviews and focus groups should greatly inform the creation and improvement of new online resources. As students are the end-users of the resources, feedback on their experiences is vital to the development process. To involve students in this step of development, Dunham et al. (2018) found that both individual interviews and focus groups were beneficial, and students should ideally be involved in both types of sessions. As an example, McKagan et al. (2008) described the involvement of students in the development of the PhET suite of simulation tools for teaching physics.

An important trait of assessment models like the above is their cyclical nature: users are involved throughout the development process, which continues through several iterations as needed. If in practice students only interact with resources deemed finished by the developers, there is no opportunity to incorporate feedback and insight from potential end-users in the development process.

*Focus Groups and Think-Aloud Sessions for Resource Development*

Focus groups have long been a useful tool for the collection of qualitative data (see, e.g., Robinson, 2020). Their interactive nature provides an advantage over individual interviews, since researchers can observe participants sharing ideas and opinions, and even debating each other (Duggleby, 2005). Including such sessions in the development process can be greatly beneficial to improving an educational resource (Dunham et al., 2018).

Similar to focus groups, "think-aloud" sessions are a method of studying a user's mental processes by asking them to speak aloud as they work on a task (Lewis, 1982). Think-aloud sessions are a common usability evaluation method, used at various steps of a project's design and development process (e.g., Nørgaard & Hornbæk, 2006; Damico & Baildon, 2007). One example of their usefulness is in the development of the PhET suite of simulation tools for teaching physics (McKagan et al., 2008). The researchers found that these sessions were very insightful and greatly improved the development process for the simulation tools, allowing the researchers to discover interface, pedagogical and programming issues in the simulation tools. Bokhove (2011) also conducted think-aloud sessions for assessing digital algebra tools with twelfth grade students, and found that the feedback was very useful for improving the digital tools.

Finally, it is important to note that there are limitations of in-person sessions. These include practical limitations, such as scheduling and attendance, as well as limitations in the types of responses a researcher may solicit from the participants, such as sharing of sensitive topics. Both of these points are discussed in further detail below. These considerations, in addition to the conversion of many activities to an online format in light of the COVID-19 pandemic, have resulted in a higher interest in online focus groups in recent research involving learners.

*Online vs. In-Person Focus Groups*

Much can be gleaned from a comparison of online versus in-person focus groups and think-aloud sessions. The use of online focus groups as a method of feedback and qualitative data collection has been increasing, even prior to COVID-19. In particular, the comparison to in-person focus groups has been of some interest. Many researchers have reported that online focus groups are at least as effective in gathering information as in-person focus groups, and even have some advantages.

Researchers have found that online focus groups work well for discussing sensitive topics. Woodyatt et al. (2016) conducted both online and in-person focus group discussions on a sensitive topic (the experience of intimate partner violence). The researchers found that sensitive topics were discussed more candidly in the online setting than in the in-person setting, and included more sharing of in-depth stories. In a similar vein, Dodds & Hess (2020) compared online group interviews on a sensitive topic (youth alcohol consumption and family communication) conducted during COVID-19 lockdown with face-to-face interviews conducted prior to COVID-19. They list several benefits to the online setting, including being comfortable, non-intrusive, and safe; and having easy communication and convenient set-up. They also found that the participants were more engaged in the discussions.

Dodds & Hess's (2020) finding on the degree of participant engagement echoes another finding from Woodyatt et al. (2016). The researchers found that the online focus group discussions resulted in a larger word count while being shorter in time than the in-person focus group discussions. Therefore, one might infer that no information was lost in the online setting. Similarly, Richard et al. (2021) conducted an experimental study to compare the diversity of responses in online focus groups to the diversity in in-person focus groups, in brainstorming sustainable practices in the hospitality industry. They found that both types of groups generated an approximately equal number of unique ideas. A thematic analysis revealed a high degree of overlap in themes between the two groups: 77% of themes occurred in both treatment groups, and these overlapping themes represented 91% of all keywords generated across both groups. These findings on idea diversity in the online setting are promising.

Finally, online focus groups are efficient and easy to set up. Matthews et al. (2018) conducted video-enabled online focus groups for a qualitative research study to explore the factors influencing the national implementation of advanced practitioner radiation therapists within Australia. They found that the online focus groups were easier to attend due to the geographically dispersed nature of their participants. Robinson (2020) added several further advantages to online focus groups: inclusivity, for people who would otherwise have to make arrangements to physically attend the focus group; accessibility, since people with physical or speech disabilities may find it easier to participate; and lowered cost for the researchers.

All of these advantages suggest that working online in groups can be a very valuable tool for obtaining feedback from participants in a variety of settings. In this paper, we present the results of online student groups on developing new educational resources. Previous work in the education literature has used (mostly) in-person focus groups and similar group sessions to gauge student opinion. However, in this work we are primarily interested in online sessions that provide actionable feedback, which we can then use to improve resource development.

*Structure of This Paper*

The next section of this paper describes in more detail the resources that we tested as well as the online focus group methodology, including the recruitment process and special considerations for the online setting. Then, we present our focus group findings. Finally, a brief discussion is given in the final section.

The student sessions discussed throughout the rest of this paper included elements of both focus groups and think-aloud sessions. We refer to our sessions as *focus groups* throughout the rest of this paper. However, it is important to keep in mind that they did include a "think-aloud" aspect, and that this feature provided many useful insights to us throughout the sessions.

**Methodology**

In order to assess our in-development resources, we carried out a series of online student focus groups. To present the methodology of our study, we first describe the resources that we assessed and then explain the recruitment process and logistics of the focus groups.

*Resources*

One of the aims of this study was to improve recently-developed (but not yet finalized) resources for the instruction of introductory and intermediate statistics by observing students engaging with them. The resources we assessed were developed for use at the University of British Columbia (UBC), a large public university in western Canada. These resources and more are freely available on StatSpace (2021); instructors and students from other academic institutions can use and adapt them.

The resources that we assessed in this study focus on the concept of *power of a statistical test* (see, e.g., Agresti & Franklin, 2007, Ch. 9.6). This concept is important in practice in many areas. For instance, a researcher's explicit specification of power determines the appropriate sample size for a study. Indeed, journals often require authors to explicitly describe their power considerations (see, e.g., the JAMA Network, 2022). However, many students struggle with the concept of power since its definition involves technical terms and complex sentence structures. Interactive resources are a promising way of aiding students' understanding of challenging concepts like this.

Each of the resources that we assessed is composed of two parts: an interactive, online web app and an accompanying activity sheet. We developed the web apps in R Shiny (R Core Team, 2020; Chang et al., 2021; Iannone et al., 2020). R is a popular, free programming language for statistical computing, and Shiny is a set of libraries for building interactive web apps in R. Figure 1 in the Findings section shows a screenshot of one of the web apps used in this study. We highlight that using the apps does not require knowledge of R programming. The computer code used to generate the web apps is also freely available on GitHub (Diluvi, 2021). The interactive web apps consist of tabs with various settings that the user can modify, as well as a graphical display. The settings are quantities relevant to the concept of power of a statistical test. For one of the web apps, e.g., these quantities correspond to the average and variance of the population and the desired specificity of the test. The graphical display within that web app contains two plots that change according to the selection of these settings. The plots reflect how different characteristics of the population and of the test affect the statistical power (and how the power relates to the specificity). This is motivated by the fact that, even though the power of a hypothesis test is a difficult concept to grasp in the abstract, it is easy to represent visually.

The inclusion of an accompanying activity sheet is motivated by previous research: Lane and Peres (2006) suggest that interactive simulations are most effective if students' interaction with them is structured, i.e., if students are directed by questions instead of freely interacting with the web app. In our case, we developed PDF files with structured questions. The activity sheets also include a preamble with learning objectives and prerequisite knowledge, which is useful for instructors that wish to use the resources. Each web app has a single accompanying activity, but we highlight that it is possible to design more activities—or adapt existing ones—if instructors want their students to interact with different aspects of the web app.

So far, we have developed two resources focusing on the concept of power. These resources differ only in their setting: one has observations from a single population and the other one from two populations. In this paper, we focus on the single-population resource, which we tested in all but one of the focus groups.

*Focus Groups Organization*

We carried out online student focus group sessions to obtain student feedback on our resources in development. To have a more diverse pool of student volunteers, we considered two target populations. The first consisted of students taking an introductory statistics course. These students had taken at least one college-level course before, but no statistics courses. The second target population consisted of students taking an intermediate statistics course. In this case, students had taken both a university-level calculus course and at least one other university-level statistics course. We highlight that the concept of power is relevant in—and in fact part of the curriculum of—both of these courses.

We contacted the instructors of the two courses (introductory and intermediate) and they agreed to send an email invitation to all of the students enrolled in the Summer 2020 and the Fall 2020 offerings of these classes There are usually between 150 and 300 students enrolled in each, depending on the class and the term. We aimed to have focus groups with between four and seven students in them, but we could not invite all students who expressed interest in the email invitation since some of the resulting groups would be too small due to students' limited availability. Hence, we scheduled focus groups with 22 students in total, 9 from the introductory class and 13 from the intermediate class. These 22 students were split between five focus groups that took place between August and November, 2020. The focus groups took place on Zoom (Zoom Video Communications, Inc., 2021), the preferred videoconference software of the university where the study took place. This ensured that both students and researchers were familiar with the software prior to the session. We obtained approval from the university's Research Ethics Board before the recruitment process. Each student attendee had to sign a consent form before the session, and we gave them a $20CAD bookstore gift card after the focus group.

During the focus groups, our interactions with students were only a brief introduction where we explained the logistics of the focus group and a ten-minute discussion at the end of the session where students could volunteer feedback. During the rest of the focus group, the students worked on the activity sheets unprompted by us, which mimics the setting in which they would encounter these resources after deployment. To encourage students to work as a team, we split them into groups of two or three students and assigned them roles. One student shared their screen with the web app on it and was in charge of manipulating the settings in it; a second student read the questions on the activity sheet, led the discussion, and was in charge of formulating an answer to each question. When three students were present, the third student carried out this last task. We obtained feedback from students both by observing them engaging with the web app and the activity and by asking them during the final discussion about their experience.

**Findings**

We split the focus group discussion findings into three themes: the online setting, i.e., the impact that holding online focus groups has on student engagement in resource development; student learning, i.e., findings related to how the resources impacted conceptual learning; and user experience, i.e., which design features of the web apps and activity sheet had an impact (either positive or negative) on students' engagement with the material.

To complement some of our findings, we include verbatim transcriptions – obtained manually by us, i.e., without the use of a transcription software – of parts of the focus groups. In those instances, we have assigned fictitious names to students in the transcription. We also draw comparisons with a series of (unpublished) in-person focus groups that we carried out between 2016 and 2018, before the COVID-19 pandemic.

*The Online Setting*

The entire process, from sending a call for volunteers to conducting the focus groups, was done completely (and seamlessly) online.

We first note that the time investment from students was less when compared with in-person focus groups. To participate in an in-person focus group, students have to budget not only for the time required to attend the focus group, but also for the time needed to travel to and from the focus group location, including commuting to and within the university campus. In contrast, the time commitment of attending a one-hour focus group was exactly one hour, even if students did not have a class that day. We also observed that students were just as likely to attend an online focus group: in our previous in-person focus groups, 35 out of 47 students (74.5%) who expressed interest showed up to the focus group, while for the online focus groups that figure was 22 out of 29 (75.9%). This difference is not significant ($p>0.1$), suggesting a similar attendance rate across the two focus group settings.

Researchers also benefited from the online setting since they typically only go to campus on certain days of the week. However, they could easily connect to online focus groups even if they were off campus. These time savings came at no cost to the quality of the information obtained from the focus group. We observed that students were as willing to share information about the resource and activity during the online focus groups as they were during in-person focus groups.

Some of the advantages of the online setting might be less significant as in-person instruction becomes the norm again. For example, if students are on campus, they might need to find a place where they can connect to the focus group, which might be harder than just participating from home. However, we believe that most of the advantages from the online setting will remain even as we return to in-person instruction, partly because students and instructors alike will still be familiar with the online setting. An example of this is the continued prevalence of hybrid or fully online work in professional and academic settings. Indeed, since the return to in-person instruction, there has been increased interest in blended learning courses.

*Student Learning Experience*

Students in the introductory statistics course managed to work through the whole activity sheet, but during the discussion at the end they said that they had not grasped the main concept in the end. This was also evident when observing their interaction with the web apps, and we realized that the activity was designed in such a way that students could move forward without understanding the concepts. For example, consider the following question: *What is the power of the t-test when the population mean is equal to 1? How is this value represented in the bottom plot?* in relation to the screenshot of the one-sample t-test web app in Figure 1.

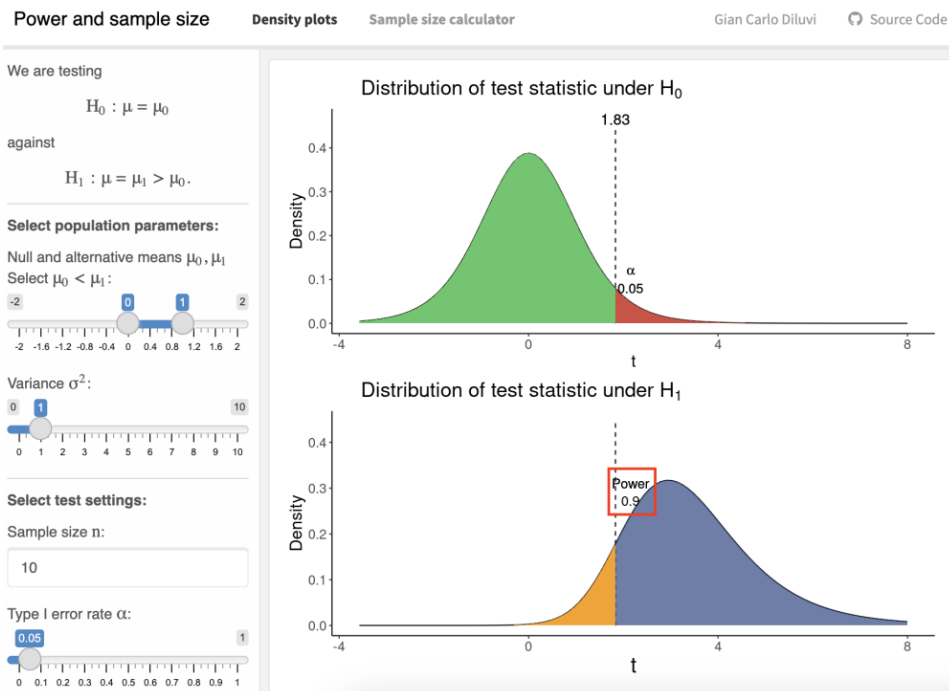

*Figure 1: Screenshot of a t-test web app. Power highlighted in square in bottom plot.*

The power of the t-test is calculated automatically by the web app and displayed on top of the blue area, and is equal to 0.9. This value, which we highlighted in a square in the bottom plot of Figure 1, is equal to the blue area in that same plot (the entire area in the bottom plot is equal to 1). Students should know this if they understand the definition of power given in the preamble of the activity sheet (or alternatively through the material of the course they were enrolled in).

However, consider the following interaction between Aaron (A) and Barbara (B), both students (with fictitious names) in the introductory statistics course that participated in the first focus group:

> A: [Reading hint] So the power is displayed on top of the blue area of the lower curve of the panel.
>
> B: Oh, so it's 0.9, okay, 0.9 and, uhm, it should be like, again, the area under the curve. Not sure how, but yeah, the power should be 0.9, and...
>
> A: Okay, okay, and...
>
> B: Is it the area under the blue curve, or under the orange...
>
> A: On top, the power is displayed on top, yeah, I think it's just...
>
> B: Yeah, it's displayed on top but, like, I'm not sure of which part is showing the, uh, power in general.
>
> A: Uh.
>
> B: Like, is it the purple-ish area or the orange area? I'm not sure. I know the value is 0.9, but I'm not sure of which part it is. Like, under which curve, which side of the curve can we see the power.
>
> A: Okay, uh, next question then.

In this interaction, Barbara was actually very close to getting the correct answer: the power is *equal* to the blue area under the curve. However, Aaron was satisfied with just finding the *value* of the power—which was displayed on top of the blue area, and was 0.9. We believe that Aaron did not grasp the visual significance of the concept of power. After agreeing with Barbara, Aaron suggests moving on to the next question—a behaviour that we frequently found when a student did not seem to understand a question. Based on how these and other students engaged with the resource, we believe that most of the students in the introductory statistics class did not understand the concept of power even after working through the activity sheet. There are many reasons why this might be the case, such as the little experience that these students have with undergraduate statistics in general. This is not something we could address by modifying the resource. However, we also observed that students in the first focus group tended to skim the preamble of the activity—where we included the definition of power—before attempting the questions. We hypothesized that this might be another reason why students were not grasping the concept of power. To address this, we modified the activity sheet by enclosing the definition of power in a box, thereby highlighting it in case students skimmed through the preamble in subsequent focus groups.

Previous research (Lane and Peres, 2006) suggests that instruction based on structured simulations—such as our combination of web app and activity sheet—might be more effective by encouraging students to *predict* the effect of modifying the web app *before* modifying it, a method called "query first." Following this line of thought, we asked students to guess the outcome of modifying certain settings in the web app; then we asked them to modify the settings; and finally, we asked them to criticize their own initial guess in light of their observations. We noticed that students who did not have a good grasp of the concept they were being asked about tended to guess incorrectly and thus were forced to reflect why. This ultimately resulted in students being more likely to reflect deeply and try again, which is a key step towards successfully understanding the core concepts. During one of the discussions, students volunteered that they liked being asked to think about what they expect will happen if they interact with the web app in a specific way.

Students in the intermediate statistics course seemed to connect the concepts with the mechanical steps more easily than those in the introductory course, and they could even develop and explore their own questions while following the flow of the activity sheet. To give an example, we consider three important concepts in statistics—variance, effect size, and power—and their relationship in the context of hypothesis testing. Succinctly, if the variance increases, the effect size decreases, and the power of the test decreases; see Krzywinski and Altman (2013).

At some point in the activity, students were asked a series of questions to help them understand this relationship. After interacting with the web app, but before reaching that series of questions, Cameron (C), Danielle (D), and Fred (F)—all students (with fictitious names) in the intermediate statistics course—started thinking about the relationship between these concepts, and how they could test whether what they thought was correct:

> C: So, you're saying, you're saying, as effect size [increases, it] increases power.
>
> D: Power should decrease.
>
> C: Hmm. I guess we could test it out. (Laughs.) Yeah, I guess we could, we could try. Yeah, let's say, hmm, let's say you make, hmm
>
> F: variance bigger...
>
> C: variance bigger, yeah.
>
> F: It should be decreasing it [the power of the test].
>
> C: Power, right?
>
> D: (Increases variance in the web app) The effect size went down and power went...
>
> F: and power went down.
>
> C: Oh, went down? Oh, so same, same [relationship] then...

In this interaction, Cameron decides to manipulate the resource to determine whether decreasing the effect size (accomplished by increasing the variance) would decrease or increase the power. Cameron then realizes that their original intuition ("as effect size [increases, it] increases power") was correct, as opposed to the inversely proportional relationship suggested by Danielle ("power should decrease").

Based on how students engaged with the material, we realized that the web app and activity were more effective for students that had previously had more exposure to the topics, as was showcased by the interaction between Cameron, Danielle, and Fred. But this assessment also pointed to the need of developing more introductory resources.

*User Experience*

We noticed that students were comfortable offering advice and constructive criticism, especially during the final discussion part at the end of each session. Whenever the students agreed that a certain change would have benefited their understanding or improved their learning experience, we incorporated their feedback into the resource and activity sheet before the next focus group.

For example, a group of students mentioned that the activity sheet seemed to have far too much text. Some even asked to have additional formulae. When first developing the resource, we intentionally included as few formulae as possible to prevent students from relying too much on them: our aim was for students to engage with the concepts more so than any specific equation. However, upon listening to students' comments, we realized that judiciously including some formulae could help students relate the concepts with the mechanical steps that they are taking. Indeed, previous research (Kuo et al.; 2013) suggests that advanced students *blend* their conceptual understanding with formulae in the problem-solving process. After adding a few key formulae to the web app, we found that students leveraged these equations while engaging with the web app to increase their conceptual understanding. For example, a formula giving the relationship between effect size and variance helped students to understand that increasing one of these quantities would decrease the other. They then further verified this new knowledge by setting different values of the variance in the web app and observing how the effect size changed.

Students did not always agree on whether a change in the web app or activity sheet would benefit them, however. For example, in one focus group some students voiced their concern of round sliders in the web app, mentioning that they would prefer input-boxes for text. Other students joined the discussion and argued otherwise. Ultimately, the whole group concluded that both were fine and it depended on personal preference. Based on this student feedback, no changes were made to the sliders. Other features of the web app were (accidentally) tested. One student, for example, confirmed that the colour palette used in the plots was colour-blind friendly. We purposely designed it like so, but had not planned to properly test it.

**Conclusion**

In this work, we developed a novel online methodology to assess resources in development via student focus groups. We tested our methodology by observing students engaging with two newly-developed resources—all in an online setting. This allowed us to obtain invaluable feedback both on the resources and on the online methodology for assessing them.

In terms of the resources, we found that advanced students engaged with the core concepts considerably more than less experienced students. This was evidenced by students' interaction with the web apps: students in the introductory course worked through the activity often without fully grasping the main concept; advanced students, on the other hand, independently hypothesized how different concepts were related to each other and then used the web app to test their understanding.

As for the online setting of our focus groups, we found that it did not come with a compromise in the quality of the information we obtained when compared to previous in-person focus groups. Furthermore, online focus groups proved to be easier to organize and resulted in time savings for both students and researchers. As schools go back to in-person instruction, modifications to the logistics of online focus groups may be necessary. However, we believe that the benefits of the online setting—especially considering that we found no obvious drawbacks for either researchers or learners—make online student focus groups a promising way of improving resources in development through student engagement.

We highlight that the information one can obtain from a one-hour focus group with few students is inherently limited. Future work can explore ways of obtaining feedback from more students, for example by incorporating a resource in development as part of a course's curriculum. Another promising area for future research is comparing resources that have benefitted from student feedback with resources that have not. As before, it might be necessary to carry out a long-term study to obtain conclusive results.